\documentclass[structabstract]{aa}  
\usepackage{graphicx}
\usepackage{longtable}
\usepackage{amsmath}
\usepackage{subfig}
\usepackage{txfonts}
\usepackage{natbib}
\begin{document}
   \title{Observing a column-dependent $\zeta$ in dense interstellar sources: the case of the Horsehead Nebula}

   \titlerunning{$\zeta (N_H)$ in Dense Sources: The Case of the Horsehead Nebula}

   \author{P. B. Rimmer\inst{1}
          \and
          E. Herbst\inst{1,2} 
          \and 
          O. Morata\inst{3}
          \and
          E. Roueff\inst{4}
          }

   \institute{Department of Physics, The Ohio State University, 
              Columbus OH 43210 USA
              \email{pbrimmer@mps.ohio-state.edu}
         \and
             Departments of Chemistry, Astronomy, \& Physics, University of Virginia,
             Charlottesville, VA 22904 USA 
         \and
				 Institute of Astronomy and Astrophysics, Academia Sinica, P.O. Box 
			    23-141, Taipei 10617, Taiwan
         \and
             Observatoire de Paris-Meudon, 
             LUTH, 5 place Jules Janssen, 92190 Meudon, France
             }

   \date{Received; Accepted}   
 
  \abstract
   {Observations  of small carbon-bearing molecules such as CCH, $\rm{C_4H}$, c\textendash$\rm{C_3H_2}$, and HCO in the Horsehead Nebula have shown these species to have higher abundances towards the edge of the source than towards the center.}
   { Given the determination of a wide range of values for  $\zeta$ (s$^{-1}$),  the total ionization rate of hydrogen atoms,  and the proposal of a column-dependent $\zeta(N_H)$, where $N_H$ is the total column of hydrogen nuclei, we desire to determine if the effects of $\zeta(N_H)$ in a single object with spatial variation can be observable. We chose the Horsehead Nebula because of its geometry and high density.}
   {We model the Horsehead Nebula as a  near edge-on photon dominated region (PDR), using several choices for $\zeta$, both constant and as a function of column. The column-dependent $\zeta$ functions are determined by a Monte Carlo model of cosmic ray penetration, using a steep power-law spectrum and accounting for ionization and magnetic field effects. We consider a case with low-metal elemental abundances as well as a sulfur-rich case. }
   {We show that use of a column-dependent $\zeta(N_H)$ of $5 \times 10^{-15}$ s$^{-1}$ at the surface and $7.5 \times 10^{-16}$ s$^{-1}$ at $A_{\rm V} = 10$ on balance improves agreement between measured and theoretical molecular abundances, compared with  constant  values of $\zeta$.}


   \maketitle
%

\section{Introduction}

Ion-neutral reactions are the most important driving processes for gas-phase chemistry. Therefore it is important to understand the mechanisms by which chemical species in the interstellar medium become ionized, in order to have a more accurate picture of the chemistry in various interstellar sources.
Near the edge of dense clouds and throughout diffuse clouds, UV photons can provide a powerful ionizing force upon the medium, especially if there is a sufficiently strong source of radiation nearby. These photons do not penetrate very far into dense clouds, decreasing exponentially with the column density. Other ionizing agents, like X-rays, will penetrate farther into dense clouds, but deep within the object, high-energy ($\gtrsim 100$~MeV) cosmic rays are the dominant ionizing force.

The recent detection of unexpectedly large abundances of H$_{3}^{+}$, however,  in an assortment of diffuse clouds has raised the old question as to whether the high ionization rate needed is caused by  a high flux of cosmic rays of $< 1$ GeV \citep{McCall2003,Indriolo2007}.  Such low energy cosmic rays would not be expected to penetrate deeply into dense clouds, so that a column-dependent ionization rate due to cosmic rays might exist in denser sources.   This question is best explored by examining the influence of cosmic ray ionization at different depths into a single object, and the Horsehead Nebula is an ideal candidate for such an investigation.

The Horsehead Nebula, also called Barnard 33, is a dark nebula of size about $5'$ in the bright nebula IC434. It is illuminated by $\rm{\sigma~Ori}$ from a distance of about $30'$ \citep{Anthony1982}. The radiation field incident on the cloud is most commonly taken to be $\chi = 60$ in Draine units \citep{Draine1978,Habart2005}, and the geometry of the cloud is described as nearly ``edge-on'', meaning that the line between $\rm{\sigma~Ori}$ and the Horsehead Nebula is nearly perpendicular to the line of sight. This makes the Horsehead Nebula ideal for observing column-dependent variables in a single source. It has an ambient magnetic field of $< 6$ $\mu$G \citep{Zaritsky1986} and a steep density gradient ranging from  $10^2$ to $10^5$ cm$^{-3}$, and contains a pre-stellar core as well as at least one other dense region near the ``throat'' that will be able to be studied in greater detail by the Atacama Large Millimeter Array (ALMA) \citep{Ward-Thompson2006}.

High abundances of small carbon-bearing molecules were observed by \citet{Teyssier2004} and by \citet{Pety2005}, with higher abundances of certain molecules ($\rm{CCH}$, c\textendash$\rm{C_3H_2}$, $\rm{C_4H}$) observed near the edge than at the center. 
This led Pety et al. to posit that polycyclic aromatic hydrocarbons (PAH's) near the edge of the cloud are being destroyed by incident radiation, and that the products of their destruction are these small hydrocarbons. 
A number of other molecules have been detected, including the ions HCO$^{+}$ and HOC$^{+}$ \citep{Goic2009},  the carbon-bearing neutrals 
$\rm{HCO}$ \citep{Gerin2009}, l\textendash$\rm{C_3H}$, and c\textendash$\rm{C_3H}$ \citep{Teyssier2004}, and the sulfur-bearing species $\rm{CS}$ and $\rm{HCS^+}$ \citep{Goic2006}, although, except for HCO$^{+}$, little information of their column dependence is available.

Chemical modeling of the Horsehead Nebula was discussed by \citet{Winnewisser1993}. \citet{Teyssier2004} provided the first detailed chemical PDR model of the Horsehead Nebula, using the Meudon PDR code \citep{LeBourlot1993,LePetit2002}.  A year later, \citet{Pety2005} modeled the Horsehead Nebula with the same code, comparing the results with observations at three different lines of sight, and incorporating PAH's into the model. \citet{Habart2005}  determined a column-dependent temperature via thermal balance. None of these models is able to account for the high abundances of small hydrocarbons at the edge, or the $\rm{HC_3N}$ abundance.   

Deuterium fractionation of HCO$^{+}$ ($[\rm{DCO^+}]/[\rm{HCO^+}] \sim 0.02$) has  been observed in the Cloud region at $A_{\rm{V}} \approx 10$, and used to constrain its temperature to about $20$ K \citep{Pety2007}. Neutral atomic oxygen has also been detected \citep{Goic2009b}, with hopes for Herschel's heightened resolution to provide abundances of atomic oxygen for different regions in the cloud.

The effect of a higher sulfur abundance was considered by \citet{Goic2006}, using the Meudon code \citep{LePetit2006}. \citet{Pety2007} also used this code to better understand deuterium fractionation at the Horsehead edge. The abundance of the negative ion $\rm{C_6H^-}$ was calculated by \citet{Millar2007}, although negative ions have not yet been observed in this region. \citet{Morata2008} developed a time-dependent PDR code, and first applied it to the Horsehead Nebula, with mixed success. This is the code we make use of in this paper, in tandem with the Meudon PDR code, which we use to determine some of the physical conditions. 

\citet{Compiegne2007} and \citet{Goic2009} have performed some recent modeling of the Horsehead region; \citet{Compiegne2007} explored the dust emission. \citet{Goic2009} self-consistently modeled the observed spatial distribution and line intensities with detailed depth-dependent predictions  coupled with a nonlocal radiative transfer calculation for ${\rm H^{13}CO^+}$, ${\rm DCO^+}$ and ${\rm HOC^+}$. They compared their model results with the \citet{Gerin2009} observations of $\rm{HCO^+}$ in order to constrain the electron fraction. Goicoechea et al. determined a very steep relative electron abundance of $n_e/n_H \sim 10^{-4} - 10^{-8}$ (where $n_H = n(\text{H}) + 2n(\rm{H_2})$) at $A_{\rm{V}} \approx 0.6-2.0$ from the cloud edge, based on a faint emission line attributed to $\rm{HCO^+}$ near the edge.

In this paper, we report our investigation on the effect of a column-dependent cosmic ray ionization rate $\zeta(N_H)$ on model results for molecular abundances and their spatial variation in the Horsehead Nebula. This is offered as a partial explanation of the high abundances of small hydrocarbons at the edge of the Horsehead nebula.  In Section~\ref{sec:zeta}, we discuss the determination of three different $\zeta(N_H)$ functions,  including a discussion of the role played by the magnetic field.   In Section \ref{sec:model}, we provide a detailed description of the PDR model used, and compare our calculated abundances with observational values using two different sets of elemental abundances.   We also provide predicted abundances for observable species.   In Section \ref{sec:conclusion}, we discuss the implications of these results, and a better determination of $\zeta(N_H)$ from single sources  after the advent of ALMA.

\section{The Determination of  $\zeta(N_H)$}
\label{sec:zeta}

The cosmic ray ionization of the interstellar medium is caused primarily by relativistic protons, alpha particles, and electrons. This ionization rate, labeled $\zeta$,  is typically represented as a per-second rate at which cosmic rays ionize atomic hydrogen.  Given the process
\begin{equation*}
\mathrm{H + CR \rightarrow H^+} + e^- + \mathrm{CR},
\end{equation*}
where CR represents ionizing cosmic rays, $\zeta$ is defined by the kinetic equation
\begin{equation*}
\dfrac{d[\mathrm{H^+}]}{dt} = \zeta [\mathrm{H}],
\end{equation*}
where the brackets signify concentration.  The ionization rate of other species, such as $\rm{H_2}$ and He, is usually determined in chemical networks by multiplying $\zeta$ by a constant. Even near the edge of dense clouds, the majority of hydrogen is molecular in nature, so it is important to note that, to a good approximation, $\zeta_{H2} \approx 2 \zeta$ \citep{Glassgold1974}.

 In the last decade, results from diffuse sources \citep{McCall2003,LePetit2004,Indriolo2007}, including recent observations with Herschel \citep{Gerin2010,Neufeld2010}, have most often indicated that in these environments $\zeta$ is more than an order of magnitude higher than the generally accepted value of $10^{-17}$ s$^{-1}$.  Earlier values for $\zeta$  ranging from $10^{-17} - 10^{-15}$ s$^{-1}$ had been proposed \citep{Spitzer1968,Hartquist1979,Dalgarno2006}. Table \ref{tab:cr} contains a limited historical overview of some of the values of $\zeta$ utilized in previous models.  These actually refer to molecular rather than atomic hydrogen.

\begin{table}
\caption{Some values of $\zeta_{H2}$ used in previous models}
\label{tab:cr}    
\centering                          
\begin{tabular}{c c}      
\hline\hline                 
 $\zeta_{H2}$ ($\times 10^{-17}$ s$^{-1}$) & Source \\  
\hline                        
	100 &	\citet{Solomon1971}	\\
	1 & \citet{Herbst1973}	\\
	$3 - 2000$ & \citet{Hartquist1978}	\\
	$10 - 100$ & \citet{McCall2003}	\\
	25 & \citet{LePetit2004}	\\
	100 & \citet{Goto2008}	\\
	$6 - 24$ & \citet{Neufeld2010}	\\
	5000 & \citet{Gupta2010}	\\
\hline                                   
\end{tabular}
\end{table}

The observations indicating a high $\zeta$, along with this wide range of values, led us to initiate a calculation of  column-dependent functions of  $\zeta$. At the same time, \citet{Padovani2009} undertook similar calculations.  They used the ionization and energy loss cross sections for collisions between cosmic rays and atomic and molecular hydrogen \citep{Cravens1978} as well as Helium to follow the flux-spectra of cosmic rays through a cloud and, from the flux spectra as a function of position, obtained the column-dependent cosmic-ray ionization rate for a number of initial flux-spectra.  Here we report  similar calculations but with a Monte Carlo approach in which we also include magnetic field effects.

\subsection{Initial Spectrum}
\label{subsec:spectrum}

We begin by considering the form of the initial cosmic ray flux-spectrum $j(E)$ (cm$^{-2}$ s$^{-1}$ sr$^{-1}$ GeV$^{-1}$ per nucleon),  as a function of energy. The spectrum has only been directly observed within our solar system, where the solar wind would have depleted the low energy cosmic rays \citep{Parker1958}.

Different cosmic ray spectra have been proposed based on assumptions about the origin of the cosmic rays. Supernova shocks are currently the favored explanation for the origin of cosmic rays \citep{Biermann2010,Axford1981}. The spectrum due to the supernova blast alone imposes a low-energy cutoff at about $100 \; \rm{MeV}$ because of energy loss due to debris and strong magnetic field effects \citep{Hayakawa1961,Ip1985}. It is suspected that shocks in the debris may re-accelerate some of the thermalized cosmic rays \citep{Ip1985,Indriolo2009}. 

Shock models favor a steep power law for low-energy cosmic rays, with a new cutoff at 1 MeV, below which most cosmic rays would again lose a significant fraction of their energy into the debris, and would either be reabsorbed into the remnant, or would travel too slowly to propagate throughout the galaxy. Alternate theories for cosmic ray acceleration exist, but these also predict similar spectra for low-energy cosmic rays \citep{Butt2009}.

 Comparison between measurements of the cosmic ray flux and theoretical cosmic ray spectra have been very useful. Basic statistics, ``leaky-box'' models, convection methods, and Monte Carlo methods have been applied to better constrain cosmic ray spectra, often by examining the elemental composition of the cosmic rays themselves. \citet{Strong2007} contains an excellent review of these methods. \citet{Webber1998} incorporated the newest results from Voyager into their Monte Carlo model, in order to determine the low energy spectrum better. 

 Nevertheless, Voyager is still in a region where solar winds have a substantial effect. In fact, the farther the Voyager satellite travels, the steeper the low energy spectrum becomes \citep{Webber1998}.
Indeed, as recently as \citet{Putze2011}, statistical, Monte Carlo, and ``leaky box'' models have been unable to constrain the low energy cosmic ray spectrum, due to a lack of direct measurement of low energy cosmic ray protons outside the solar influence.

\citet{Indriolo2009} list many of the proposed cosmic ray spectra. We consider three representative spectra \citep{Hayakawa1961,Spitzer1968,Nath1994}, which are shown in Figure \ref{fig:zeta}. These three spectra span the range of low energy cosmic rays.  The spectrum from \citet{Spitzer1968} is based on solar system measurements of the low energy cosmic ray flux, and contains the minimum low-energy cosmic ray spectrum. \citet{Nath1994} assume that the power-law for the cosmic ray spectrum at $1$ GeV continues down until a hard cut-off at $1$ MeV. Theirs is the highest published estimate of the low energy cosmic ray flux. We chose to use these three spectra in order to provide the full range of impact that different low energy cosmic ray flux spectra have on the ionization rate.
The spectrum of \citet{Nath1994} increases the most steeply towards lower energies,  that of \cite{Spitzer1968}  actually decreases towards lower energies, and that of \cite{Hayakawa1961} lies in the 
middle.

\subsection{Cross Sections}
\label{subsec:sigma}

We calculate the loss of energy by considering $10^{4}$ cosmic ray protons with energies, $E$ (in eV unless otherwise noted), distributed according to the three spectra selected above. The particles stream into a cloud of a number density $n$ (cm$^{-3}$). At each distance increment, the particles are each assigned a random number, which is compared with the probability of an ionizing or other inelastic collision over an incremental distance, determined by cross-sections, $\sigma$ (cm$^2$). For ionizing collisions by protons we use the form of $\sigma_i$ from \citet{Spitzer1968}. The cross section (cm$^{2}$) for ionization of a hydrogen atom as a function of $E$ and the rest-energy of the proton ($E_P$) is given by
\begin{align}
\sigma_{i,\text{H}} =& 7.63 \times 10^{-20} \left(1 - \left(\frac{E_P}{E + E_P}\right)\right)^{-1} \notag \\
&+ 1.23 \times 10^{-20} \log\left(\left(\frac{E + E_P}{E_P}\right)^2 - 1\right) \notag \\
&- 5.29 \times 10^{-21} \label{eqn:sigma}.
\end{align}
The first term is dominant for ``low'' energies ($500$ keV $< E < 50$ MeV), so for $E < 50$ MeV, $\sigma_{i,\text{H}} \propto 1/E$, down to $E \approx 500$ keV, when Equation (\ref{eqn:sigma}) ceases to be accurate. This cross-section is also used below for determining the cosmic ray ionization rate of atomic hydrogen. For molecular hydrogen, we simply multiply the cross section by a factor of 2.

Inelastic collisions are considered for atomic and molecular hydrogen only, and we use the cross-sections from \citet{Cravens1975}, accounting for rotational, vibrational and electronic excitation as well as dissociation reactions.
For the ionization of helium, the differential  cross-section from \citet{Dalgarno1999} is integrated to yield a total cross section:
\begin{equation}
\sigma_{i,\text{He}} = 1.5 \epsilon_0 A(E),
\label{eqn:helium}
\end{equation}
where $A(E)$ ($\propto 1/E$ for $E \lesssim 100$ MeV; $\propto \log(E)$ for $E \gtrsim 1$ GeV) and $\epsilon_0$ are parameters fit to the measurements of \citet{Shah1987}.

\subsection{Energy Loss}
\label{subsec:eloss}

The energy loss calculation assumes a column great enough that the cosmic rays will collide with gaseous atoms and molecules many times. Since our model is one-dimensional, we do not consider the effects of elastic collisions on the exclusion of low energy cosmic rays from molecular clouds.

Because there are many collisions, we are justified in utilizing the average energy lost by cosmic ray in an ionizing collision, $\overline{W}$. This is equal to the ionization energy plus the average energy of the ejected electron. For molecular hydrogen, this is determined from the differential cross-section by \citet{Cravens1978,Dalgarno1999} to be
\begin{equation}
\overline{W} \; (\text{eV}) = 7.92 E^{0.082} + 4.76,
\label{eqn:eloss}
\end{equation}
where $E$ (eV)  is the energy of the cosmic ray before the ionizing event. The energy losses from other types of inelastic collisions with molecular hydrogen, as well as ionizing and other inelastic collisions with $\rm{H}$ and $\rm{He}$ are taken from the detailed forms in \citet{Cravens1975}.

This energy loss is subtracted from the initial energy of the cosmic ray, and becomes the new energy. At each increment, a new flux-spectrum, $j(E,N_H)$, is calculated, and new random numbers are assigned to the cosmic rays.  Because of the energy-dependence of the $\sigma$ functions, lower energy cosmic rays have more ionizing collisions. In the case of  the spectrum of \citet{Nath1994}, cosmic rays with $E < 50$ MeV contribute $99\%$ to the value of $\zeta$ (see Section~\ref{subsec:ionization}). To complicate matters, however, there is energy loss from magnetic effects in addition to the loss from collisions.  Magnetic energy loss is assigned based on interactions with Alfv\'{e}n waves, as discussed below, using a static magnetic field of 3 $\mu$Gauss. 

\subsection{Magnetic Field Effects}
\label{subsec:magnetic}

Magnetic fields play an important role in the transport of cosmic
rays. The Lorentz force is the largest magnetic force acting on cosmic
rays, and affects energy loss by increasing the path length cosmic
rays travel, as they spiral along the magnetic field lines. This
resulting increase in path length is not, however,  the primary source of energy
loss. Rather, the dominant magnetic field effect on cosmic rays is due to
irregularities in the magnetic field.

Because of the neutralization of low-energy cosmic rays, there will be
far fewer cosmic rays at the center of the cloud than at the edge.
Since cosmic rays are overwhelmingly positively charged, these losses
introduce a charge imbalance in the cloud. Electrons are attracted to
the edge, and their motion generates magnetic field irregularities
moving from the center to the edge of the cloud with velocity
$\upsilon_A = B (4\pi\rho)^{-1/2}$. These irregularities, called
Alfv\'{e}n waves, are the dominant source of energy loss, as discussed  
in \citet{Skilling1976}. 
 \citet{Hartquist1978,Hartquist1979} first applied the work of \citet{Skilling1976} to calculate cosmic ray ionization rates, and proposed different values of $\zeta$, depending on the object. 

 Following \citet{Skilling1976}, we determine the charge imbalance using the Monte Carlo simulation
with $B = 0$, and consider it in terms of a characteristic column
density, $\lambda(E)$ (cm$^{-2}$), determined by the simulation, at which the number of cosmic rays
will be depleted by a factor of $e$. This means that 
$N_H$ must be
$\gtrsim \lambda(E)$ for cosmic rays of energy $E$ to be significantly
affected by magnetic field irregularities. This function will appear later in the analysis.

Alfv\'{e}n waves are driven by the charge imbalance, and are damped by the
friction between ions and the surrounding gas, as discussed by \citet{McIvor1975} in terms of the collision rate $\Gamma$ (s$^{-1}$) between
ions and neutrals \citep{Dalgarno1968}. The larger $\Gamma$ is, the
less effect the waves have. The static magnetic field enhances the
damping by absorbing smaller irregularities. However, the larger static
magnetic field also increases $\upsilon_A$, and thus the frequency of
collisions between the cosmic rays and the Alfv\'{e}n waves.

This mechanism for cosmic ray energy loss by Alfv\'{e}n wave effects
is important for $N_H < 10^{24}$ cm$^{-2}$ when $B \lesssim 6$ mG and
$n_H \lesssim 10^9$ cm$^{-3}$. At a given column density, cosmic ray energy is substantially affected by Alfv\'{e}n waves for energies less than the energy $E_0$. The static magnetic field outside denser regions is assumed to be much smaller than the field inside these regions.   Because the difference between the magnetic field inside and outside the cloud significantly dampens the Alfv\'{e}n waves for mid to high energy cosmic rays, $E_0$ cannot be greater than $50$ MeV \citep{Cesarsky1978}. $E_0$ is dependent on various physical parameters of the source in question. For typical cold and dense interstellar conditions, $n(\rm{HI}) = 1$ cm$^{-3}$, $n_H = 10^4$ cm$^{-3}$, $T = 20$ K, and $B = 3 \; \mu$G.  Under these conditions, the use of  $j_0(E)$ from \citet{Nath1994} leads to $E_0 = 1$ MeV at $N_H = 10^{19}$ cm$^{-2}$ , while for  $N_H > 10^{21}$ cm$^{-2}$, $E_0 = 50$ MeV. 

 Integrating over energies up to this cutoff value, we can obtain
the magnetohydrodynamic solution for $j_{\rm{IC}}(E,N_H)$, the ``In-Cloud'' cosmic ray flux-spectrum at a given $N_H$, to be \citep{Skilling1976}
\begin{align}
j_{\rm{IC}}(E < E_0,N_H) =& \dfrac{\lambda(E)}{E} \Bigg[\dfrac{E_0
j(E_0,N_H)}{\lambda(E_0)} \notag \\
& + \dfrac{2\upsilon_A}{N_H}\int_{E'=E}^{E_0}
\dfrac{\alpha j(E',N_H)}{\upsilon(E')} dE' \notag \\
& + \dfrac{U_M \Gamma}{\pi^2
m\upsilon_A \Omega_0 N_H}
\ln\Bigg(\dfrac{\gamma_0^2-1}{\gamma^2-1}\Bigg)\Bigg]; \\
j_{\rm{IC}}(E > E_0,N_H) =& j(E,N_H) 
\label{eqn:solution}
\end{align}
In  this expression, $j(E,N_H)$ is the spectrum determined using the Monte
Carlo simulation in the absence of magnetic field effects, 
the magnetic energy density $U_M = B^2/2\mu_0$ (erg/cm$^{-3}$),
 $\Omega_0$ is the gyromagnetic frequency (s$^{-1}$), the Compton-Getting
factor $\alpha$ \citep{Gleeson1968} is
\begin{equation*}
\alpha = -\dfrac{10}{9} \dfrac{E}{j_0(E)} \dfrac{\partial j_0}{\partial E},
\end{equation*}
where $j_0$ is the initial cosmic ray
flux-spectrum,
$\gamma = (1-\upsilon^2/c^2)^{-1/2}$ and $\gamma_0 =
(1-\upsilon_0^2/c^2)^{-1/2}$ where $\upsilon_0$ is the velocity of a
cosmic ray of energy $E_0$.

Given a steep initial $j_0(E)$, the approximate effect of the Alfv\'{e}n waves
and Lorentz Force is to shift the cosmic ray spectrum, and thus the
ionization rate (see next section), from 
$\zeta(N_H)$ to $ \zeta(5 N_H)$,  so that the ionization rate decreases more strongly with column. 

For cosmic ray flux-spectra that are not very steep ($m < 2$ for $j
\propto E^{-m}$), the shift is less extreme. Of course, for the full
description of the relationship of $\zeta$ to the column density,
Equation (\ref{eqn:solution}) must be calculated for $E < E_0$.

\subsection{The Column-Dependent Ionization Rate}
\label{subsec:ionization}

The value of $\zeta(N_H)$ is calculated by integrating the product of
 the flux-spectrum from eq.~(\ref{eqn:solution}), $j_{\rm{IC}}(E,N_H)$, and $\sigma_{i,\text{H}}(E)$ from eq.~(\ref{eqn:sigma}), as a function of ``depth'' $N_H$ into a cloud, with various correction factors:
\begin{equation}
\zeta(N_H) = 1.8 \times \frac{5}{3} \times \int_0^\infty 4 \pi \sigma_{i,\text{H}}(E) j_{\rm IC}(E,N_H) dE. \label{eqn:zeta}
\end{equation}
The factor of $5/3$  \citep[][]{Spitzer1968,Dalgarno1999} takes into account the additional ionization caused by secondary electrons, while the factor of 1.8 accounts for ionization due to $\alpha$ particles ($\rm{He^{+2}}$). These particles are the second most important source of ionizing cosmic rays ($\zeta_{\alpha} \approx 0.8 \zeta_p$). By comparison, relativistic electrons, the third most important ionizing source, have little effect:  $\zeta_e \sim \zeta_p / 100$.

\begin{figure}
   \centering
   \includegraphics[width=9cm]{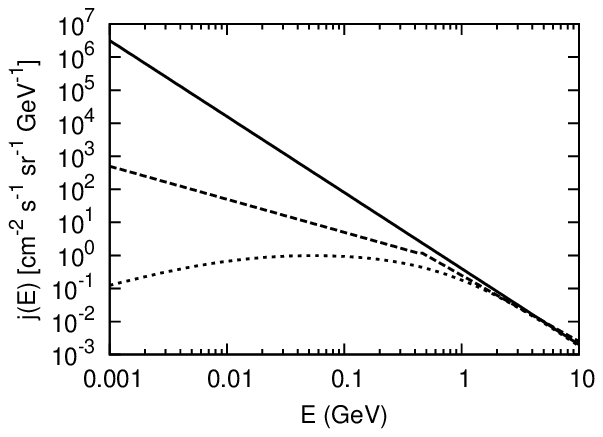}
      \caption{Three different cosmic ray flux spectra, taken from \citet{Hayakawa1961} (dashed line), 
      \citet{Spitzer1968} (dotted line), and \citet{Nath1994} (solid line).}
         \label{fig:zeta}
   \end{figure}

Three different functions for  $\zeta(N_H)$ have been calculated, based on the cosmic ray flux-spectra in Figure \ref{fig:zeta}, which are chosen to be widely divergent below $E \approx 500$ MeV to account for the uncertainty in the low-energy region \citep{Hayakawa1961,Spitzer1968,Nath1994}. Analytical expressions for $\zeta(N_H)$, used in the models below, are:
\begin{align}
\zeta_{H,\text{Hayakawa}} & = \dfrac{5 \times 10^4}{N_H} + 10^{-17} \; \text{s$^{-1}$}, \\
\zeta_{H,\text{Nath}} & = 0.002(N_H)^{-0.6} + 10^{-17} \; \text{s$^{-1}$},
\end{align}
and are valid for $10^{24}$ cm$^{-2}$ $\gtrsim N_H \gtrapprox 5 \times 10^{19}$ cm$^{-2}$. These analytical expressions do not seem to change significantly for $100$ cm$^{-3}$ $ < n < 10^6$ cm$^{-3}$ and $5$ K $ < T < 1000$ K, beyond which the effects of the density and temperature on magnetic field effects becomes significant.
 
\begin{figure}
   \centering
   \includegraphics[width=9cm]{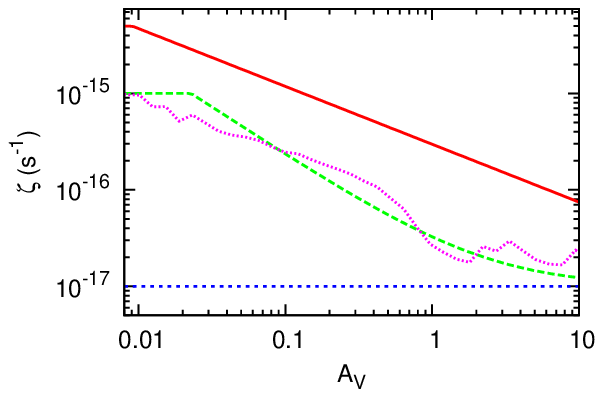}
      \caption{The results of the one-dimensional Monte Carlo model for $\zeta$ described in Section \ref{sec:zeta} in terms of $A_{\rm{V}}$. The solid red, dashed green, and dotted blue lines derive from the flux-spectra of \citet{Nath1994},  \citet{Hayakawa1961}, and \citet{Spitzer1968}, respectively. These lines fit the averaged result of dozens of iterations of the Monte Carlo model. The results from a single Monte Carlo run using the flux-spectrum of \citet{Hayakawa1961} are included (pink dotted) in order to show error. 
              }
         \label{fig:zAv}
   \end{figure}

The results are depicted in Figure \ref{fig:zAv}, in terms of the visual extinction between the cloud and the UV source ($A_{\rm V}$). We determine $A_{\rm V} \approx 4.3 \times 10^{-22} N_H$, using $Q$ efficiencies from \citet{Laor1993} with a grain distribution (in terms of the ``radius'' of the grain, $a$) of $n \propto a^{-3.5}$ with $r_{\rm min} = 5$ nm and $r_{\rm max} = 1$ $\mu$m. With these assumptions, the analytical expressions for $\zeta(A_{\rm V})$ are:
\begin{align}
\zeta_{H,\text{Hayakawa}} & = \dfrac{2.2 \times 10^{-17}}{A_{\rm V}} + 10^{-17} \; \text{s$^{-1}$}, \\
\zeta_{H,\text{Nath}} & = 3.05 \times 10^{-16} (A_{\rm V})^{-0.6} + 10^{-17} \; \text{s$^{-1}$}.
\end{align}
These expressions are later referred to as ``mid-range'' and '``high-range'' values, respectively. 

The wide range of the ionization rate demonstrates the importance of low-energy cosmic rays, especially at low $N_H$ or $A_{\rm V}$.  Other calculations of $\zeta(N_H)$ have been performed, either for high column densities ($>10^{24}$ cm$^{-2}$) where low energy cosmic rays do not penetrate \citep{Umebayashi1981,Finocchi1997}, without consideration of the magnetic field \citep{Padovani2009}, or in regions where there are no ionization losses \citep{Padoan2005}. Recently, \citet{Padovani2011} have incorporated the effect of magnetic mirroring, whereas we have treated the effects of Alfv\'{e}n waves on cosmic ray streaming. The results in this paper suggest that Alfv\'{e}n waves may have a more substantial effect on $\zeta$, with factor of $\sim$10 impact on $\zeta$ at certain $N_H$ for Alfv\'{e}n waves versus a factor of $\sim$2-4 impact on $\zeta$ from magnetic mirroring. Ultimately, a robust magnetohydrodynamics simulation of cosmic ray transport would be necessary to determine what magnetic field effects have the most significant impact on cosmic ray penetration.

In our study below, we determine the  effect of four different functions for $\zeta$ on the chemistry in the Horsehead Nebula.  We consider the $\zeta(N_H)$ functions based on flux spectra from \citet{Nath1994} and \citet{Hayakawa1961}, as well as constant  values for $\zeta$ of $10^{-15}$ s$^{-1}$, and $10^{-17}$ s$^{-1}$, the latter of which is effectively the $\zeta(N_H)$ derived from the spectrum of  \citet{Spitzer1968}. 

\section{Modeling the Horsehead Nebula}
\label{sec:model}

We have used the PDR model of \citet{Morata2008}  with the OSU  03/2008  gas-phase network.\footnote{http://www.physics.ohio-state.edu/$\sim$eric/research.html} This network is a purely gas-phase one that treats the PDR as a semi-infinite series of slabs with the radiation source impinging on one edge. It does not account for freeze-out or any surface chemistry, aside from a simple approximation for H$_2$ formation on grains and selected ion recombination processes.  Radiative transfer and self-shielding of $\rm{H_2}$ and $\rm{CO}$ \citep{Draine1996,Lee1996} are calculated in progression, starting with the slab at the edge. The chemistry is solved with a time-dependent gas-phase kinetics model for each slab.
This model, like our model for cosmic rays, is one-dimensional (1D). Because cosmic rays are thought to stream in from all sides, the effects of the geometry are mostly lost in this model. However, even with cosmic rays streaming in from all angles, low energy cosmic rays will dominate at the edge, and will be absent from the center. The average value of $\zeta$ at a slab near the edge will be close to the value determined from the 1D Monte Carlo model. Because the majority of slabs near the center will not have low-energy cosmic rays, the average $\zeta$ near the center also be close to the 1D value. Thus the average value of $\zeta$ at different slabs of the cloud will be close to the 1D values we use for $\zeta$ found in Figure \ref{fig:zAv}.

Following \citet{Pety2005}, we compare, when possible, observations with model results for three regions at different optical extinctions ($A_{\rm{V}}$) from the edge of the Horsehead PDR. These are the IR-edge ($A_{\rm{V}} = 1.56 \pm 0.73$), IR-Peak ($A_{\rm{V}} = 4.55 \pm 1.7$), and the Cloud ($A_{\rm{V}} = 11.7 \pm 4.1$). The error bars in $A_{\rm V}$ are based both on the beam size of the observations and uncertainty in the density profile of the cloud, as discussed in the next section. We determine the error in fractional abundance by taking the ratio between the observed column density of the species and the error in that column density, both from \citet{Pety2005}.

\subsection{Physical Conditions and Initial Chemical Abundances}
\label{sec:initial}

\begin{figure}
   \centering
   \includegraphics[width=9cm]{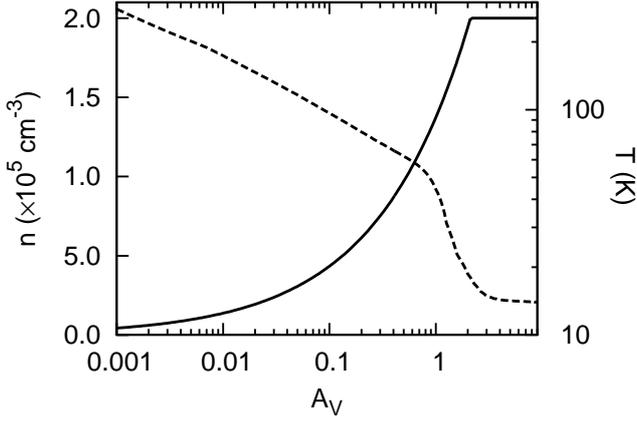}
      \caption{The temperature (dashed line) and density (solid line) profiles as functions of visual extinction with $\zeta_{H,\text{\rm Nath}}$. The density profile is in the form of \citet{Habart2005}, our equation (\ref{eqn:density}), with $\beta = 1$. The temperature is from thermal balance \citep{LePetit2006}. At $A_{\rm V} = 10$, $\zeta \approx 10^{-16}$ s$^{-1}$, which raises the temperature by $\approx 4$ K at the center compared to a $\zeta$ of $10^{-17}$ s$^{-1}$.
              }
         \label{fig:nT}
   \end{figure}

The density profiles used are taken from \citet{Habart2005}. The temperature profile is calculated from thermal balance \citep{LePetit2006}. Cosmic rays heat the interstellar medium through the thermalization of secondary electrons and photons,  \citep{Field1969,Glassgold1973}. Thermal heating by cosmic rays begins to dominate at $A_{\rm{V}} > 3$, but the thermal impact of different cosmic ray ionization rates is not very significant until $\zeta > 10^{-16}$ s$^{-1}$. Since even the highest $\zeta(N_H)$ drops to $\approx 10^{-16}$ s$^{-1}$ at the Cloud region, the temperature difference here between the high $\zeta(N_H)$ and $\zeta = 10^{-17}$ s$^{-1}$ is only about 4 K. The density and temperature profiles are shown in Fig.~\ref{fig:nT}.

 The gas density increases with spatial distance into the nebula as a power law with an exponent $\beta$ \citep{Habart2005}, which in terms of column density can be written as:
\begin{equation}
n_H(N_H) =
\begin{cases}
n_{H,0}\Bigg[\dfrac{(\beta+1)N_H}{x_0 n_{H,0}}\Bigg]^{\beta/(\beta+1)} & N_H \leq N_{H,0}\\
n_{H,0} & N_H > N_{H,0},
\end{cases}
\label{eqn:density}
\end{equation}
where $\beta \geq 1$ is a dimensionless constant used to parameterize the steepness of the number density, $n_{H,0} = 2 \times 10^5$ cm$^{-3}$, $x_0 = 0.02$ pc is a length scale, and $N_{H,0} =  (1+\beta)^{-1} 1.23 \times 10^{22}$ cm$^{-2}$ is the column density at a depth of $x_0$. For our analysis, we show the results for $\beta = 1$, and discuss results for both $\beta = 1$ and $\beta = 4$. The steeper density gradient impacts the UV photon flux and the resulting thermal balance. There are different total densities for the IR-edge and IR-peak regions. The difference in UV penetration, temperature and density at different values of $A_{\rm V}$ noticeably impacts the chemistry. The cosmic ray ionization, however, is not significantly altered by the density gradient, because for the ranges of density of $100$ to $10^5$ cm$^{-3}$, $\zeta$ is column-dependent, but not density dependent.

Other densities and density profiles have been proposed. \citet{Pety2005} used several uniform number densities and profiles, while \citet{Goic2009} proposed a slowly changing piecewise function for the density, with three sections instead of two, reaching $2 \times 10^5$ cm$^{-3}$ at $A_{\rm{V}} \approx 5$ instead of $A_{\rm{V}} \approx 1.0$, as used here. Until the number density is better determined,  significant uncertainties in the extinction at a given angular depth will persist.

The UV radiation field impinging on the Horsehead surface has been a topic of much discussion and uncertainty \citep{Anthony1982,Zhou1993,Abergel2003}. Values of $\chi = 30$ to $\chi = 100$ in Draine units \citep{Draine1978} have been proposed. We use  $\chi = 60$, because this is the most commonly used value for the Horsehead PDR. The external UV field is important to the chemistry only for the IR-edge. For the IR-peak and the Cloud regions, cosmic rays are the primary ionizing and photochemical agent.

The initial chemical abundances used for the Horsehead PDR \citep{Lee1996b,Morata2008} are listed in Table \ref{tab:ab} and represent abundances for a dark cloud prior to the onset of a nearby star.  These abundances comprise  observed values for small (less-than-six-atom) species in TMC-1, as well as calculated early-time values from \citet{Smith2004}  for atoms and small molecules that have not been observed, based on so-called ``low-metal'' elemental abundances.

In addition to these initial abundances, we also investigated cases with much higher elemental abundances of sulfur, based primarily on the analysis of CS and HCS$^{+}$ by  \citet{Goic2006}, who place the total elemental sulfur abundance with respect to $n_{\rm H}$ at $3.5 \times 10^{-6}$. On the other hand, \citet{Teyssier2004} used a value of $[\rm{S}] \sim 10^{-7}$, similar to the low-metal value used in this part of the paper. To determine the effect of raising the sulfur abundance, we utilized elemental abundances for sulfur, relative to hydrogen, of $10^{-6}$ and $10^{-5}$, starting primarily from the neutral atomic form.

The abundances are calculated from time $t = 0$ to steady state ($t = 5 \times 10^6$ yr).  Since the age of the Horsehead Nebula is not well-determined, values from $10^{4} - 10^{6}$ yr have been considered \citep{Morata2008}. We focus only on the time of $10^5$ yr, because in general the calculated results are closest to observational values at this time. We also use this time because it is a reasonable age for a molecular cloud, given its size and velocity gradient \citep[see][]{Pound2003}. Time-dependence was investigated by 
\citet{Morata2008} albeit with a different density profile from what is used here.   They found that at times between 10$^{5}$ yr and steady-state, the abundances of carbon chain species in the Cloud region become sharply lower, as is found in standard cold dark clouds.  They also investigated times as early as 10$^{4}$ yr, at which time the abundance profiles are flatter. Our calculations for carbon chain species have reached steady state by $10^4$ years for $A_{\rm V} < 5$. For $A_{\rm V} > 5$, our calculations confirm their findings.

In Figures~\ref{fig:c2h} to~\ref{fig:cs},   we show the calculated abundances of various molecules as continuous functions of visual extinction with observed values in boxes to delineate the uncertainties in both abundance and $A_{\rm V}$. The calculated abundances are plotted with two fixed values of $\zeta$: $10^{-17}$ s$^{-1}$ and $10^{-15}$ s$^{-1}$, as well as with two column-dependent ionization rates depicted in Figure \ref{fig:zAv}: the  mid-range $\zeta(N_H)$ (dashed green line), and the high-range $\zeta(N_H)$ (solid red line).  The fixed value of $\zeta = 10^{-17}$ s$^{-1}$ is equivalent to the lowest-range $\zeta(N_H)$ in Figure \ref{fig:zAv}, which utilizes only high-energy protons.  Neither of the two fixed values for $\zeta$ is likely to be physically reasonable; the low value can pertain to the inner Cloud region but is less likely to pertain to a region near the edge, where at least some low-energy cosmic rays exist, while the high value is more likely to pertain only to the edge of the PDR.    Unless specified, the low elemental abundance of sulfur  is utilized.

\begin{table}
\caption{Initial fractional abundances with respect to $n_H$}             
\label{tab:ab}      
\centering                          
\begin{tabular}{c c c c}        
\hline\hline                 
Species & $f(X)$\tablefootmark{1} & Species & $f(X)$\tablefootmark{1} \\    
\hline                        
   $\rm{H_2}$ & 0.5 & $\rm{C_2H}$ & $1.0 \times 10^{-8}$ \\
   $\rm{H}$ & $7.5 \times 10^{-5}$ & $\rm{CO_2}$ & $1.3 \times 10^{-8}$ \\
   $\rm{He}$ & 0.14 & $\rm{H_2O}$ & $3.5 \times 10^{-8}$ \\
   $\rm{C}$ & $2.8 \times 10^{-8}$ & $\rm{HCN}$ & $1.0 \times 10^{-8}$ \\
   $\rm{O}$ & $1.0 \times 10^{-4}$ & $\rm{HNC}$ & $1.0 \times 10^{-8}$ \\ 
   $\rm{N}$ & $1.3 \times 10^{-5}$ & $\rm{NH_3}$ & $1.0 \times 10^{-8}$ \\
   $\rm{S}$ & $7.2 \times 10^{-8}$ \tablefootmark{2} & $\rm{SO_2}$ & $5.0 \times 10^{-10}$ \\
   $\rm{Si}$ & $7.8 \times 10^{-9}$ & $\rm{C_3H}$ & $5.0 \times 10^{-9}$ \\
   $\rm{Cl}$ & $4.0 \times 10^{-9}$ & $\rm{C_4H}$ & $4.5 \times 10^{-8}$ \\ 
   $\rm{Fe}$ & $3.9 \times 10^{-10}$ & $\rm{c-C_3H_2}$ & $5.0 \times 10^{-9}$ \\
   $\rm{Mg}$ & $1.9 \times 10^{-9}$ & $\rm{HC_3N}$ & $1.0 \times 10^{-8}$ \\
   $\rm{Na}$ & $4.7 \times 10^{-10}$ & $\rm{C^+}$ & $4.7 \times 10^{-9}$ \\
   $\rm{P}$ & $3.0 \times 10^{-9}$ & $\rm{H^+}$ & $4.2 \times 10^{-10}$ \\ 
   $\rm{CH}$ & $1.0 \times 10^{-8}$ & $\rm{He^+}$ & $3.5 \times 10^{-10}$ \\
   $\rm{CN}$ & $2.5 \times 10^{-9}$ & $\rm{Fe^+}$ & $2.6 \times 10^{-9}$ \\
   $\rm{CO}$ & $7.3 \times 10^{-5}$ & $\rm{Mg^+}$ & $5.1 \times 10^{-9}$ \\
   $\rm{CS}$ & $2.0 \times 10^{-9}$ & $\rm{Na^+}$ & $1.5 \times 10^{-9}$ \\ 
   $\rm{N_2}$ & $4.2 \times 10^{-6}$ & $\rm{S^+}$ & $1.2 \times 10^{-9}$ \\
   $\rm{NO}$ & $1.5 \times 10^{-8}$ & $\rm{Si^+}$ & $2.5 \times 10^{-10}$ \\
   $\rm{O_2}$ & $8.1 \times 10^{-8}$ & $\rm{H_3^+}$ & $1.4 \times 10^{-9}$ \\
   $\rm{OH}$ & $1.0 \times 10^{-7}$ & $\rm{HCO^+}$ & $4.0 \times 10^{-9}$ \\
   $\rm{S_2}$ & $1.8 \times 10^{-9}$ & $\rm{HCS^+}$ & $2.0 \times 10^{-10}$ \\
   $\rm{SO}$ & $1.0 \times 10^{-9}$ & $\rm{N_2H^+}$ & $2.0 \times 10^{-10}$ \\
\hline                                   
\end{tabular}
\tablefoot{\tablefoottext{1}{$f(X)=n(X)/(n(H) + 2n(H_2))$}\\
\tablefoottext{2}{The sulfur-rich cases include $[\text{S}] = 10^{-6}$ and $10^{-5}$}.}
\end{table}

\subsection{Results: $\rm{C_2H}$, $\rm{c-C_3H_2}$ and $\rm{C_4H}$}
\label{sec:cres1}

Hydrocarbons are not direct tracers of $\zeta$; nevertheless, an enhanced $\zeta$ at the surface of the Horsehead nebula may help to explain the high abundances of these small hydrocarbons at the edge. $\rm{C_2H}$, $\rm{c-C_3H_2}$ and $\rm{C_4H}$ are formed by a complex network of reactions, linked at least partially to the cosmic ray ionization rate via several sequence of reactions based on C and C$^+$.  The sequence involving neutral atomic C starts with the reactions:
\begin{align}
\mathrm{H_2} + \mathrm{CRP} & \rightarrow \mathrm{H_2^+} + e^- + \mathrm{CRP} \\
\mathrm{H_2^+} + \mathrm{H_2} & \rightarrow \mathrm{H_3^+} + \mathrm{H} \\
\mathrm{C} + \mathrm{H_3^+} & \rightarrow \mathrm{CH^+} + \mathrm{H_2},
\end{align}
and $\rm{CH^+}$ initiates a series of chemical reactions that eventually results in $\rm{C_2H}$, $\rm{c-C_3H_2}$ and $\rm{C_4H}$ via recombination with electrons.   The C$^{+}$ ion is produced in three ways depending on physical conditions: at low extinction ($A_{\rm V} < 2.5$), it is formed principally by photoionization, and can reach a fractional abundance as high as 10$^{-4}$, whereas at high extinction ($A_{\rm V} > 4.5$) it is formed less efficiently by the reaction between He$^+$ and CO. In the middle region ($2< A_{\rm V} < 5$), secondary photons from cosmic rays form a large amount of the C$^+$. Once produced,  it can radiatively associate with H$_2$ to form the CH$_{2}^{+}$ ion, which initiates a series of reactions similar to those initiated by CH$^{+}$ \citep{Herbst2008}. Because of these alternate pathways, small hydrocarbons may not be as sensitive to $\zeta$ very close to the edge or deep within the Horsehead PDR. Regardless, our robust chemical network allows us to explore in detail the effect of a column-dependent $\zeta$ on the Horsehead Nebula.

\begin{figure*}
   \centering
   \includegraphics[width=12cm]{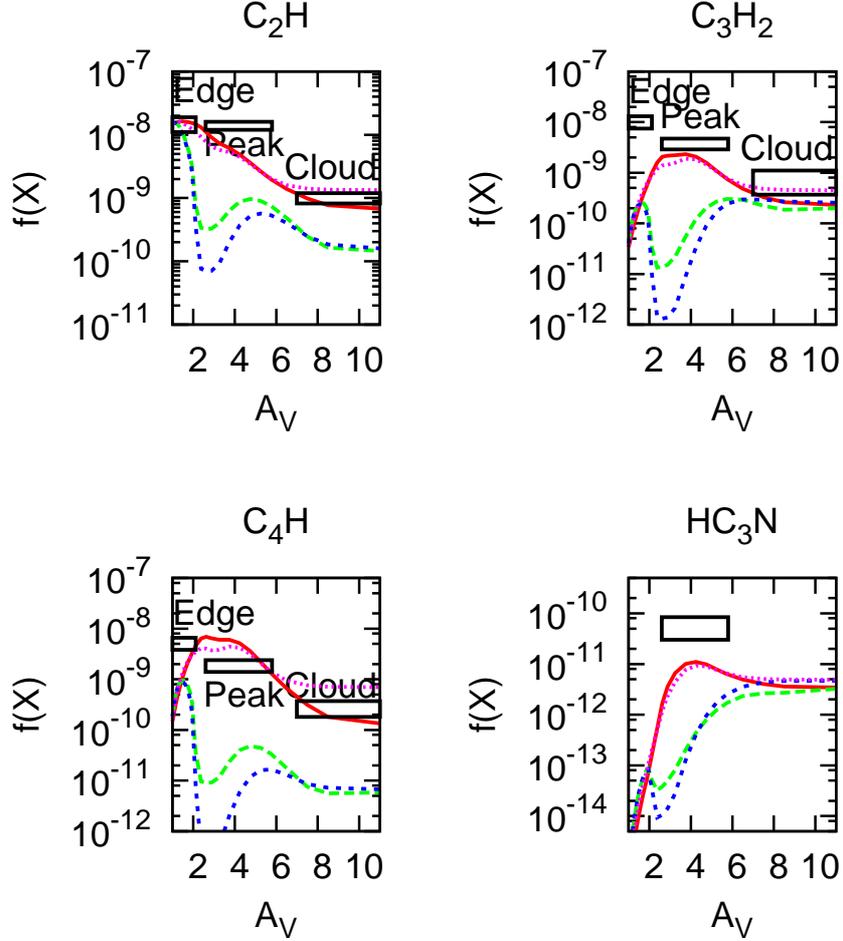}
      \caption{Fractional abundances of $\rm{C_2H}$, $\rm{c-C_3H_2}$, $\rm{C_4H}$, and $\rm{HC_3N}$ as functions of $A_{\rm{V}}$. The boxes represent observations  with error bars, and the lines are the model results for $\zeta = 10^{-17}$ s$^{-1}$ (blue dashed), $\zeta = 10^{-15}$ s$^{-1}$ (pink dotted), and, from Figure \ref{fig:zAv}, the mid-range $\zeta(N_H)$ (green dashed) and high-range $\zeta(N_H)$ (red solid).}
         \label{fig:c2h}
   \end{figure*}

The model abundances for $\rm{C_2H}$, $\rm{c-C_3H_2}$, and $\rm{C_4H}$ vs $A_{\rm V}$ can be found in Figure \ref{fig:c2h}, where observed abundances with estimated uncertainties are plotted as boxes for the three regions studied: the IR-edge, the IR-peak, and the Cloud. For $\rm{C_2H}$, our use of temperature and density profiles seems to account for the observed abundance at the IR-edge, regardless of the value of $\zeta$, probably because $\rm{C_2H}$ formation is so dependent on photon effects at the edge. The results diverge for the IR-peak and Cloud, where the high-range $\zeta(N_H)$ and  $\zeta = 10^{-15}$ s$^{-1}$ seem to do better than the other two choices of $\zeta$.   In the IR-peak, the abundances obtained with the high-range $\zeta(N_H)$ and  $\zeta = 10^{-15}$ s$^{-1}$ come within a factor of $\approx 5$ of the observed value, and are closer still for the Cloud region.  

For $\rm{c-C_3H_2}$, and for $\rm{C_4H}$,  none of the four plots comes particularly close to the observed values at the center of the IR-edge, although the curves obtained with the high-range $\zeta(N_H)$ and  $\zeta = 10^{-15}$ s$^{-1}$ graze the lower portion of the observation box for C$_{4}$H.   This discrepancy suggests that, though a high surface $\zeta$ is important, there are likely other factors that must be taken into account, such as PAH fragmentation \citep{Pety2005}. For the IR-peak region,   the high-range  $\zeta(N_H)$ and  $\zeta = 10^{-15}$ s$^{-1}$ models lead to results that graze portions of the observational boxes for both species , with the others models exhibiting much too low an abundance.  Finally, for the Cloud region, the high-range  $\zeta(N_H)$ and  $\zeta = 10^{-15}$ s$^{-1}$ models do quite well for C$_{4}$H, and c-C$_{3}$H$_{2}$. while the lower ionization models show reasonable agreement only for the latter.

It would appear that, on balance, the results obtained with the high constant $\zeta$ and the high-range column-dependent $\zeta$ are closer to observation in most instances for these three hydrocarbons.  To further distinguish between these two sets of results, we focus on the abundance ratios between  IR-peak and Cloud regions for the three carbon-chain species. The  ratios are taken at the visual extinctions where the models agree best with the observations, and are listed in Table \ref{tab:molrat}. The reason for taking these ratios is that we can better compare results between a fixed and a column-dependent ionization rate in this manner. These ratios are examined only as a way to distinguish between a constant and a column-dependent $\zeta$, and their use beyond this function is severely limited. For example, the ${\rm C_2H}$ emission attributed to the Cloud region may be from the FUV illuminated surface \citep[for an analogous example involving HCO, see][]{Gerin2009}. It is likely that the observed ratios will change and will be far better constrained when the Horsehead Nebula is explored at higher angular resolution.

For $\rm{C_2H}$ and $\rm{c-C_3H_2}$, the ratios are much closer to observation for the column-dependent $\zeta(N_H)$ than for $\zeta = 10^{-15}$ s$^{-1}$. In both of these cases, the ratios from the $\zeta(N_H)$ model are within a factor of $2$ of the observed ratios. For $\zeta = 10^{-15}$ s$^{-1}$, model ratios disagree by a factor of 5-7. In the case of $\rm{C_4H}$, the ratio from the constant $\zeta$ agrees slightly better with observations than for $\zeta(N_H)$, although the ratios of both models are close to observation. Also, examining the $\rm{C_4H}$ abundances from Figure \ref{fig:c2h}, it is evident that, within the Cloud region, the results from $\zeta(N_H)$ are much closer to observation than the results from $\zeta = 10^{-15}$ s$^{-1}$. In summary, as well as being unphysical, the results from a model with a constant $\zeta = 10^{-15}$ s$^{-1}$ do not agree as closely as the results from a model with a column-dependent $\zeta(N_H)$.  

For $\beta = 4$ at $t = 10^5$ yr, the results are not significantly changed for the IR-edge, and the model underestimates the small hydrocarbon abundances for the IR-peak region by about an order of magnitude. The reasons for this seem to be as involved as the hydrocarbon chemistry. The most significant factor is that the production of these hydrocarbons at a higher density requires a higher $\zeta$, and for $\beta = 4$, the density is much higher at the IR peak than for $\beta = 1$. Also, with the steeper density gradient, photons are more effective at ionizing and dissociating at the IR-Edge, but fall off more abruptly at higher $A_{\rm V}$. The difference in C$^+$ formation by photons between $\beta = 1$ and $\beta = 4$ density profiles is a factor of three, and only present at $A_{\rm V} < 2.5$.

It should finally be mentioned that thermal balance from photons depends somewhat on the density, and so the temperature profiles with $\beta = 1$ and $\beta = 4$ are different. At $A_{\rm V} = 0.001$, the gas-phase temperature for $\beta = 1$ is about $300$ K, where for $\beta = 4$, $T \approx 600$ K. The gas-phase temperatures for the two density profiles converge at $A_{\rm V} = 1$, and this undoubtedly has some impact on the chemistry. It should be emphasized that $\zeta(N_H)$ is similar for the steep and gradual density gradients.

\begin{table}
\caption{Abundance Ratios for Carbon-chain Species Between IR-peak and Cloud Regions}
\label{tab:molrat}    
\centering                          
\begin{tabular}{l c c c}      
\hline\hline                 
 & \multicolumn{3}{c}{IR-peak/Cloud} \\
Species & Obs. & $\zeta(N_H)$ & $10^{-15}$ s$^{-1}$\\  
\hline                        
   $\rm{C_2H}$ & 14 & 8.6 & 2.7 \\
   $c$\textendash$\rm{C_3H_2}$ & 25 & 13 & 3.3 \\
   $\rm{C_4H}$ & 5 & 8 & 4\\
\hline                                   
\end{tabular}
\tablefoot{The model results are obtained at 10$^{5}$~yr.  Those listed under $\zeta_{H}(N_{H})$ are for the high-range column-dependent $\zeta$ in Fig. \ref{fig:zAv}. Observations are from \cite{Pety2005}.\\
}
\end{table}

\subsection{Results: $\rm{HC_3N}$, $\rm{HCO^+}$, $\rm{HCO}$ and the electron fraction}
\label{sec:cres2}

Only one line of the carbon-chain species HC$_{3}$N has been detected, and this with a very large beam-size \citep{Teyssier2004}. We follow Teyssier's tabulated value for $A_{\rm V}$, and treat the emission as originating in the IR-peak, though there is  some uncertainty about the origin of this emission. The four models all under-produce the observed abundance of $\rm{HC_3N}$ by a little less than an order of magnitude or more, as can be seen in Figure \ref{fig:c2h}, with the models with the high-range $\zeta(N_{H})$ and the fixed $\zeta = 10^{-15}$ s$^{-1}$ coming closest.   

Cyanoacetylene ($\rm{HC_3N}$) is not as dependent as the other species on cosmic ray ionization for much of the range of visual extinction. Two reactions primarily lead to its formation:
\begin{eqnarray*}
\mathrm{CN} + \mathrm{C_2H_2} \rightarrow \mathrm{HC_3N} + \mathrm{H},\\
\mathrm{C_3H_2N^+} + e^{-} \rightarrow \mathrm{HC_3N} + \mathrm{H}.
\end{eqnarray*}
At the edge, the first reaction is directly related to $\zeta$ through $\rm{C_2H_2}$, but the second reaction involves $\mathrm{C_3H_2N^+}$, the formation of which is not strongly dependent on $\zeta$. In the Cloud region, the situation is reversed: $\rm{C_2H_2}$ is less dependent on $\zeta$, and $\mathrm{C_3H_2N^+}$ is then closely linked with cosmic ray ionization. Because of the two channels for $\rm{HC_3N}$ we expect less dependence on $A_{\rm V}$ except in the middle range:  $1 < A_{\rm V} < 5$. The results, shown in Figure \ref{fig:c2h}, roughly bear this out.  Interestingly, both the observed and calculated abundances for HC$_{3}$N are much lower than the initial value, which is taken from the TMC-1 abundance.  The discrepancy with the Cloud value, over three orders of magnitude, is especially large and very different from the analogous cases for the hydrocarbons in Figure~\ref{fig:c2h}.

\begin{figure*}
   \centering
   \includegraphics[width=14cm]{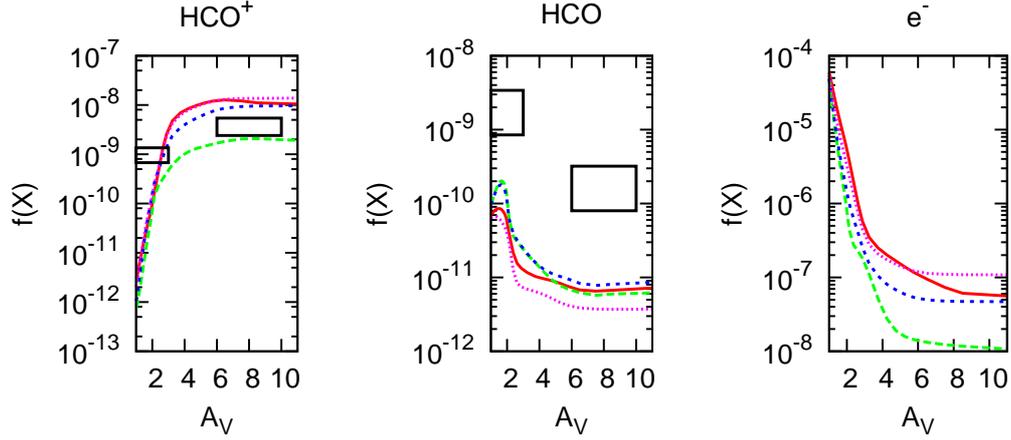}
      \caption{Relative abundances of $\rm{HCO^+}$, $\rm{HCO}$ and the ionization fraction as functions of $A_{\rm{V}}$. The boxes represent observations with error bars, and the lines are the model results for $\zeta = 10^{-17}$ s$^{-1}$ (green dashed), $\zeta = 10^{-15}$ s$^{-1}$ (pink dotted), and,  from Figure \ref{fig:zAv}, the mid-range $\zeta(N_H)$ ( blue dotted) and high-range $\zeta(N_H)$ (red solid).}
         \label{fig:hco+}
   \end{figure*}

 Figure \ref{fig:hco+} contains the observations and model results for $\rm{HCO^+}$ and HCO.  Since HCO$^{+}$ is optically thick, the carbon-13 isotopologue was used for observations.
 $\rm{H^{13}CO^+}$ was observed in emission at $\approx 40''$ from the PDR edge \citep{Gerin2009}, corresponding to an $A_{\rm{V}} \approx 10$, which is essentially the Cloud region. Following the analyses of \citet{Gerin2009} and \citet{Goic2009}, we determined the abundance of $\rm{HCO^+}$ from $\rm{H^{13}CO^+}$ by assuming $\rm{^{12}C}/\rm{^{13}C} = 60$. A faint emission feature attributed to $\rm{H^{13}CO^+}$ was also seen at $\approx 10''$ from the PDR edge, corresponding to an $A_{\rm{V}} \approx 2$ with our density profile, and so lies essentially at the IR-edge.  

In the immediate neighborhood of  $A_{\rm V}$ = 2, however, none of the models produces enough HCO$^{+}$, but the increase in abundance with increasing extinction is steep and by $A_{\rm V}=3$, all but possibly the $\zeta = 10^{-17}$ s$^{-1}$ model produce a comparable result to what is observed at the IR-edge. \citet{Goic2009} did much better fitting the $\rm{HCO^+}$ abundances at the edge by including PAH's. They also modeled profiles for $\rm{H^{13}CO^+}$ and ${\rm DCO^+}$.

 For the Cloud value, all models are in reasonable agreement with observation for HCO$^{+}$, coming within factors of 2-5 of the observed abundance.  The formation of $\rm{HCO^+}$ by cosmic rays is very direct at high extinction; in regions where UV photons cannot penetrate, it is almost solely the product of the destruction channel for protonated molecular hydrogen with carbon monoxide. At the IR-edge, however, the UV driven formation by the reactions $\rm{H_2 + CO^+}$ and $\rm{H_2O + C^+}$ dominates. In all regions, $\rm{HCO^+}$ is destroyed mainly by recombination. 

For neutral HCO, all model results are too low by an order of magnitude or more at both the IR-edge and Cloud regions,  even with the relatively fast reaction between CH$_{2}$ and O \citep[from][]{Gerin2009}.   Our results disagree with the model results from \citet{Gerin2009} and \citet{Goic2009} partly because the Meudon reaction network includes a formation mechanism absent in the OSU network, the photodissociation reaction
\begin{equation}
{\rm H_2CO} + h\nu  \rightarrow {\rm HCO + H},
\label{eqn:HCO-destruction}
\end{equation}
where $h\nu$ represents an external UV photon. This reaction is also discussed in \cite{Gerin2009}. Including this reaction enhances the HCO abundance by a factor of $5$ in the PDR,  bringing the HCO abundance within an order of magnitude of the observed value.

The ionization fraction, $f(e^-)$, is a measure of elemental abundances, ionization rate, density, and chemistry, as well as a constraint on the coupling of the magnetic field to the matter in the cloud. The ionization fraction from our models, as shown in Figure \ref{fig:hco+}, ranges from $\sim 10^{-4}$ in the PDR to $\sim 10^{-8}$ in the Cloud region. This range of fractions agrees generally with the profile in \citet[][their Figure 4]{Goic2009}. Their inferred profile for the ionization fraction would favor the mid-range $\zeta(N_H)$ from the cosmic ray flux-spectrum of  \citet{Hayakawa1961}.

For the steeper density profile with $\beta = 4$ and the high $\zeta(N_H)$, our results are somewhat different. The $\rm{HCO^+}$ abundances are not significantly changed, and the modeled $\rm{HCO}$ abundances increase by a factor of two in the IR-Edge and IR-Peak regions (at 10$^5$ yr).  Significantly, our calculated abundance of $\rm{HC_3N}$ comes into good agreement with the Cloud region observation; it is a factor of $3$ higher than the observed abundance at $t=10^5$ yr.

\subsection{Tabulated Abundances}

Calculated fractional abundances (with respect to $n_{\rm H}$) obtained with the standard elemental abundances are listed for more than twenty species in Table \ref{tab:mol}, including both observed and undetected molecules.  The calculated results are for a time of 10$^{5}$ yr and pertain to the center points of the IR-edge, IR-peak, and Cloud regions \citep{Pety2005}, for which observational results are also shown, when available. Some of the tabulated abundances, $\rm{HOC^+}$ especially, seem to be possible tracers for the cosmic ray ionization, because their fractional abundance becomes more dependent on the extinction when $\zeta$ depends on column density, than when $\zeta$ is a constant value.

In this table, we consider only the model with the  high-range  $\zeta(N_H)$, because it is evident that, at least for carbon-chain species, use of this column-dependent $\zeta$ leads generally to better agreement with observations than models with lower ionization, and it is more physical than the constant high-ionization model. Also, we do not include the case of the steeper density profile in this table. Predictions are discussed below in Section~\ref{sec:prediction}.

\begin{table}
\caption{Observations and model results  for fractional abundances at $10^5$ yr.}
\label{tab:mol}    
\centering                          
\begin{tabular}{l c c c c c c}      
\hline\hline                 
Species & \multicolumn{2}{c}{IR-edge} & \multicolumn{2}{c}{IR-peak} & \multicolumn{2}{c}{Cloud} \\
 & Obs. & Mod. & Obs. & Mod. & Obs. & Mod. \\  
\hline                        
	O $\; (10^{-5})$  &  & 11 & & 7.3&   & 5.9\\ 
   N $\; (10^{-6})$    & & 16 &  & 1.1 &   & 1.3\\ 
   CN $\; (10^{-8})$    &  & 2.3 &  & 1.0 &  & 0.2\\ 
   NO $\; (10^{-9})$   &  & 0.07 &  & 77 &   & 103\\ 
   O$_2 \; (10^{-7})$  & & $<$0.01 &  & 7.2 &   & 62\\
   OH $\; (10^{-7})$   & & 0.01 & & 1.8 &   & 1.3\\ 
   	$\rm{CO} \; (10^{-5})$ & & 6.0 & & 9.6 & & 9.9\\
	$\rm{H_2O} \; (10^{-9})$ & & 0.4 & & 170 & & 193\\
   $\rm{C_2H} \; (10^{-8})$ & 3.3 & 1.6 & 3.0 & 0.3 & 0.2 & 0.07\\
   $c$\textendash$\rm{C_3H} \; (10^{-10})$ & & 1.9 & 5.4 & 0.6 & & 1.4\\
   $l$\textendash$\rm{C_3H} \; (10^{-10})$ & & 1.0 & 2.9 & 8.0 & & 0.9\\   $c$\textendash$\rm{C_3H_2}  \; (10^{-10})$ & 13 & 2.3 & 11 & 10 & 0.4 & 2.3 \\
   $\rm{C_4H}  \; (10^{-9})$ & 9.5 & 1.3 & 3.6 & 2.0 & 0.8 & 0.1 \\
	$\rm{CH_4} \; (10^{-9})$ & & 0.03 & & 33 & & 50\\
	$\rm{C_6H} \; (10^{-11})$ & & 1.4 & & 4.1 & & 0.2\\
	$\rm{HCO} \; (10^{-10})$ & & 0.03 & 17\tablefootmark{a} & 0.1 & & 0.2\\
	$\rm{NH_3} \; (10^{-8})$ & & $<0.01$ & & 1.2 & & 2.5\\
	$\rm{HCN} \; (10^{-10})$ & & 0.8 & & 40 & & 26\\
	$\rm{HNC} \; (10^{-10})$ & & 0.9 & & 67 & & 17\\
   $\rm{HC_3N}  \; ( 10^{-11})$ & & $<0.01$ & 5.7\tablefootmark{b} & 0.7 & & 0.3 \\
   $\rm{HC_5N} \; ( 10^{-12})$ & & $<0.01$ & & 4 & & 0.13 \\
	$\rm{CH^+} \; ( 10^{-12})$ & & 5 & & 0.02 & & $<0.01$\\
	$\rm{CO^+} \; (10^{-13})$ & $\leq 5$\tablefootmark{c} & 20 & & 3.2 & & 1.3\\
   $\rm{HCO^+}  \; (10^{-9})$ & 0.9\tablefootmark{d} & 0.02 & & 10 & 3.9\tablefootmark{d} & 11\\  
   $\rm{HOC^+}  \; (10^{-12})$ & 4\tablefootmark{e} & 6 & & 73 & & 27 \\  
	$\rm{OH^+} \; (10^{-13})$ & & 4 & & 15 & & 5\\
	$\rm{H_2O^+} \; (10^{-13})$ & & 8 & & 33 & & 12\\
	$\rm{H_3O^+} \; (10^{-10})$ & & 0.1 & & 45 & & 37\\
	$\rm{CH_3^+} \; (10^{-11})$ & & 23 & & 10 & & 0.2\\
	$\rm{C_2H_4^+} \; (10^{-13})$& & 1.5 & & 19 & & 12\\
   $\rm{CS}  \; (10^{-8})$ & 1.6\tablefootmark{f} & 0.04 & 4.0\tablefootmark{f} & 0.04 & & 0.01 \\  
   $\rm{HCS^+} \; (10^{-11})$ & & 0.07 & 4.0\tablefootmark{g} & 0.08 & & 0.1 \\ 
      
\hline                                   
\end{tabular}
\tablefoot{The IR-edge is at $A_{\rm{V}} = 1.56 \pm 0.73$, the IR-peak is at $A_{\rm{V}} = 4.55 \pm 1.7$ and the Cloud region is at $A_{\rm{V}} = 11.7 \pm 4.1$. The model results are for the high column-dependent  $\zeta$ from Fig. \ref{fig:zAv}. Observations are from \cite{Pety2005} unless otherwise noted.\\
\tablefoottext{a}{$\rm{HCO}$ at $A_{\rm{V}} \approx 4$ by \cite{Gerin2009}.}\\
\tablefoottext{b}{$\rm{HC_3N}$ observed at $A_{\rm{V}} \approx 4$ by \cite{Teyssier2004}.}\\
\tablefoottext{c}{Upper limit at $A_{\rm{V}} < 2$, from \cite{Goic2009}.}\\
\tablefoottext{d}{$\rm{HCO^+}$ from $\rm{H^{13}CO^+}$ at $A_{\rm{V}} \approx 2$ and $A_{\rm{V}} \approx 10$ by \cite{Gerin2009}. The calculated abundance at $A_{\rm V} = 3$ is in close agreement with the observed edge value.}\\
\tablefoottext{e}{Observed at $A_{\rm{V}} \approx 2$ by \cite{Goic2009}.}\\
\tablefoottext{f}{For the IR-peak and IR-edge, from \cite{Teyssier2004}.}\\
\tablefoottext{g}{$\rm{HCS^+}$ observed at $A_{\rm{V}} \approx 4$ by \cite{Goic2006}.}
}
\end{table}

\subsection{The Sulfur-Rich Case}
\label{sec:sres}

We considered sulfur-bearing species, both with the standard initial abundances, and also for a sulfur-rich environment. We found that the higher the elemental sulfur (up to a relative abundance of $10^{-5}$), the closer the model matches observations for sulfur-bearing molecules. Our results and those of  \citet{Goic2006} for the chemistry and radiative transfer agree very well.

\begin{figure}
   \centering
   \includegraphics[width=9cm]{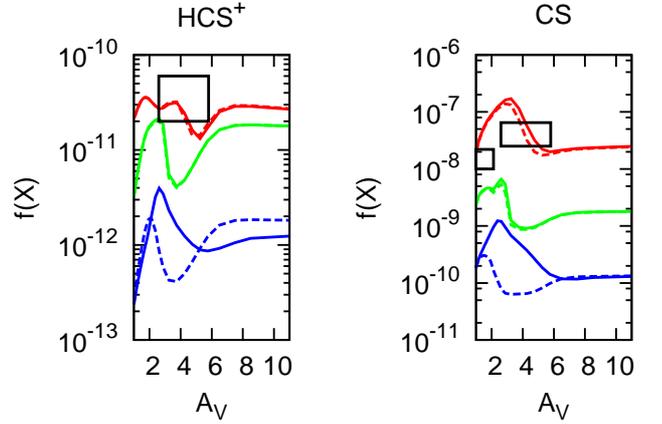}
      \caption{Relative abundances of $\rm{HCS^+}$ and $\rm{CS}$ as a function of $A_{\rm V}$. The boxes are the observations  with error bars, and the lines are the model results with $[\rm{S}] = 10^{-5}$ (red), $[\rm{S}] = 10^{-6}$ (green) and $[\rm{S}] = 7.2 \times 10^{-8}$ (blue), all using the high column-dependent $\zeta$ from Figure \ref{fig:zAv}. The solid lines use a rate for $\rm{S} + \rm{H^+} \rightarrow \rm{S^+} + \rm{H}$ of $1.3 \times 10^{-9}$ cm$^{3}$ s$^{-1}$ \citep{Prasad1980} and the dashed lines use a rate of $1 \times10^{-14}$ cm$^{3}$ s$^{-1}$
              }
   \label{fig:cs}
   \end{figure}

The results for the observed sulfur-bearing species CS and HCS$^{+}$ vs $A_{\rm V}$ at 10$^{5}$ yr  can be found in Figure \ref{fig:cs} as a function of the sulfur elemental abundance.  There are two sets of curves, depending upon the rate coefficient for the charge-exchange reaction
\begin{equation*}
\rm{S} + \rm{H^+} \rightarrow \rm{S^+} + \rm{H},
\end{equation*}
which can affect the abundances of CS and HCS$^{+}$ at low sulfur abundances.  This reaction has a listed rate coefficient of $1.3 \times 10^{-9}$ cm$^{3}$ s$^{-1}$ \citep{Prasad1980} but a more likely value of $1 \times 10^{-14}$ cm$^{3}$ s$^{-1}$ has been calculated.\footnote{This rate has been tabulated in The Controlled Fusion Atomic Data Center (http://www-cfadc.phy.ornl.gov/astro/ps/data/cx/hydrogen/rates/cti.dat).}

  The agreement attained by increasing the elemental abundance, $[\rm{S}]$, to $10^{-5}$ comes at a cost: at $10^5$ yr, all the carbon-bearing species in this scenario are reduced by up to a factor of 10 except at the IR-edge. This effect is most severe in the Cloud region. This depletion occurs in part because the high sulfur abundance destroys hydrocarbons by reactions with $\rm{S^+}$ and also with $\rm{S}$ at higher extinctions and because of the increased fractional ionization. The depletion of carbon-bearing species worsens agreement for all observed species except $\rm{HCO^+}$, which is brought to within a factor of $2$ of observation in the Cloud region.

This problem may suggest that a more realistic gas-phase sulfur elemental abundance for the Horsehead Nebula should lie somewhere around $10^{-6}$, in agreement with \citet{Goic2006}. The abundances of observed and predicted molecules  with $[\rm{S}] = 10^{-6}$ are in Table~\ref{tab:sulfur} for the same species as listed in Table~\ref{tab:mol}.  Even with this intermediate sulfur abundance, the calculated abundances of carbon chain species in particular are lowered considerably compared with the corresponding values in Table~\ref{tab:mol}, leading to worse agreement with observation.

\subsection{Some Predictions}
\label{sec:prediction}

A high column-dependent $\zeta$ brings with it  implications for chemistry in the Horsehead PDR.   This column-dependent $\zeta$ varies from $\approx 2 \times 10^{-16}$~s$^{-1}$ at the IR-edge to $\approx 7 \times 10^{-17}$~s$^{-1}$ in the Cloud region and so leads to profiles distinctive from models with fixed ionization rates, as can be seen for carbon-chain species in Figures \ref{fig:c2h} to \ref{fig:cs}.

Also, other molecules are predicted to be in amounts in principle observable, and these are listed among the species in Tables \ref{tab:mol} and \ref{tab:sulfur}. Because our $\zeta(N_H)$ produces reasonable abundances of $\rm{C_4H}$ and $\rm{HC_3N}$ in selected regions with a low elemental abundance of sulfur, we would also expect to observe, albeit with some difficulty, the more complex carbon-chains $\rm{C_6H}$ and $\rm{HC_5N}$, based on our predictions for these regions. In addition, the molecule $\rm{HCN}$ should definitely be present in observable quantities, especially in inner regions, and its isomer, $\rm{HNC}$, should also be observed with a ratio $\rm{HCN}/\rm{HNC} \approx 1$. We predict ammonia in observable quantities at $A_{\rm V} > 4$, for the low-sulfur case. 
 
 Given the observations of high amounts of the reactive molecular ions $\rm{OH^+}$ and $\rm{H_2O^+}$ in many molecular objects \citep{Gerin2010,Gupta2010}, it would be useful to consider predicted abundances of these species. Our model predictions for $\rm{OH^+}$, $\rm{H_2O^+}$ and ${\rm H_3O^+}$ in the Horsehead Nebula are contained in Tables \ref{tab:mol} and \ref{tab:sulfur}.  These predictions show low abundances for the first two ions that are rather independent of which of the three regions we consider.  The basic problem is the low abundance of atomic hydrogen except at the border of the PDR \citep{Neufeld2010}.  
Even at the IR-Edge, ${\rm H_3O^+}$ is more than an order of magnitude higher than either OH$^+$ or ${\rm H_2O^+}$, though none of these species should be sufficiently abundant to be detected. In the Cloud Region, where the electron density is at the low level of a cold dark cloud, ${\rm H_3O^+}$ is depleted rather slowly by reactions with electrons, and should achieve a high enough column to be detectable.

\begin{table}
\caption{Observations and model results for fractional abundances with $[\rm{S}] = 10^{-6}$ at $10^5$ yr.}
\label{tab:sulfur}    
\centering                          
\begin{tabular}{l c c c c c c}      
\hline\hline                 
Species & \multicolumn{2}{c}{IR-edge} & \multicolumn{2}{c}{IR-peak} & \multicolumn{2}{c}{Cloud} \\
 & Obs. & Mod. & Obs. & Mod. & Obs. & Mod. \\  
\hline                        
	O $\; (10^{-5})$  &  & 12 & & 9.9 &   & 4.9\\ 
   N $\; (10^{-6})$    & & 16 &  & 1.3 &   & 0.3\\ 
   CN $\; (10^{-8})$    &  & 1.1 &  & 0.08 &  & 0.02\\ 
   NO $\; (10^{-9})$   &  & 0.1 &  & 69 &   & 123\\ 
   O$_2 \; (10^{-7})$  & & $<$0.01 &  & 18 &   & 260\\
   OH $\; (10^{-7})$   & & 0.01 & & 1.0 &   & 1.0\\	
	$\rm{CO} \; (10^{-5})$ & & 5.5 & & 7.3 & & 7.3 \\
	$\rm{H_2O} \; (10^{-9})$ & & 0.4 & & 420 & & 390\\
   $\rm{C_2H} \; (10^{-8})$ & 3.3 & 1.1 & 3.0 & 0.3 & 0.2 & $<0.01$\\
   $c$\textendash$\rm{C_3H} \; (10^{-10})$ & & 1.8 & 5.7 & 0.2 & & 0.01 \\
   $l$\textendash$\rm{C_3H} \; (10^{-10})$ & & 1.0 & 2.9 & 0.1 & & $<$0.01\\ 
   $c$\textendash$\rm{C_3H_2}  \; (10^{-10})$ & 13 & 2 & 11 & 0.4 & 0.4 & 0.03 \\
   $\rm{C_4H}  \; (10^{-9})$ & 9.5 & 0.7 & 3.6 & 0.1 & 0.8 & $<0.01$ \\
	$\rm{CH_4} \; (10^{-9})$ & & 0.05 & & 30 & & 22\\
	$\rm{C_6H} \; (10^{-11})$ & & 1.2 & & $<0.01$ & & $<0.01$\\
	$\rm{HCO} \; (10^{-10})$ & & 0.9 & 17\tablefootmark{a} & 0.5 & & 0.06 \\
	$\rm{NH_3} \; ( 10^{-8})$ & & $<$0.01 & & $<$0.01 & & $<$0.01\\
	$\rm{HCN} \; (10^{-10})$ & & 0.6 & & 18 & & 10\\
	$\rm{HNC} \; (10^{-10})$ & & 0.9 & & 39 & & 27\\
   $\rm{HC_3N}  \; (10^{-11})$ & & $<$0.01 & 5.7\tablefootmark{b} & 0.03 & & 0.01 \\
   $\rm{HC_5N} \; (10^{-12})$ & & $<$0.01 & & $<$0.01 & & $0.01$ \\
	$\rm{CH^+} \; (10^{-12})$ & & 2.9 & & $<0.01$ & & $<0.01$\\
	$\rm{CO^+} \; ( 10^{-13})$ & $\leq 5$\tablefootmark{c} & 1.5 & & 0.7 & & 0.5 \\
   $\rm{HCO^+}  \; ( 10^{-9})$ & 0.9\tablefootmark{d} & 0.02 & & 0.7 & 3.9\tablefootmark{d} & 6.2\\    
   $\rm{HOC^+}  \; (10^{-12})$ & 4\tablefootmark{e} & 4 & & 20 & & 10 \\  
	$\rm{OH^+} \; (10^{-13})$ & & 3.1 & & 9.3 & & 3.6\\
 	$\rm{H_2O^+} \; (10^{-13})$ & & 5.3 & & 15 & & 7.4\\
	$\rm{H_3O^+} \; ( 10^{-10})$ & & 0.1 & & 50 & & 40\\
	$\rm{CH_3^+} \; (10^{-11})$ & & 20 & & 3.2 & & 0.3\\
	$\rm{C_2H_4^+} \; ( 10^{-13})$& & 2.1 & & 3.2 & & 0.8\\
   $\rm{CS}  \; (10^{-8})$ & 1.6\tablefootmark{f} & 0.5 & 4.0\tablefootmark{f} & 0.9 & & 0.2\\  
   $\rm{HCS^+} \; (10^{-11})$ & & 1.1 & 4.0\tablefootmark{g} & 1.8 & & 1.7 \\  
\hline                                   
\end{tabular}
\tablefoot{See Table~\ref{tab:mol} for footnotes.}
\end{table}

\section{Discussion}
\label{sec:conclusion}

We have modeled the Horsehead Nebula as a PDR with time-dependent gas-phase chemistry using a column-dependent cosmic ray ionization rate $\zeta(N_H)$, as well as the temperature and density profiles of \citet{Habart2005}. At a cloud age of 10$^{5}$ yr, the incorporation of a high $\zeta(N_H)$ improves agreement between model and observation for the small carbon-bearing molecules  $\rm{HCO^+}$, $\rm{HC_3N}$, $\rm{C_2H}$, $\rm{c-C_3H_2}$, and $\rm{C_4H}$ compared with a more standard constant ionization rate. With a higher abundance of elemental sulfur than our standard value, the results for small sulfur-bearing species are improved, but at the expense of our calculated values for carbon-chain species. There are also predictions of abundances and profiles for other species, some not yet observed in the Horsehead Nebula, which should be in principle observable, including HCN, HNC, NH$_{3}$, $\rm{C_6H}$, $\rm{HC_5N}$, and $\rm{H_3O^+}$.  Some of these predictions are strongly affected, however, by an increase in the assumed sulfur elemental abundance.

Our results for c-C$_{3}$H$_{2}$ and C$_{4}$H (but not for C$_{2}$H) also indicate  that the fracturing of PAH's may play an important role in the production of these molecules towards the edge of the PDR, but our model does not incorporate the effects of PAH's. Strong aromatic emission, observed by \citet{Compiegne2007}, poses some problems, however, for the hypothesis that PAH fracturing is the source of small hydrocarbons.  These authors claim a high concentration of neutral PAH's in the HII region, which suggests that PAH's may endure the radiation at the IR-edge, instead of breaking apart into the observed hydrocarbons.

The detailed form of the calculated abundance profiles in Figures \ref{fig:c2h} through \ref{fig:cs} cannot be observed because observations up to the present lack sufficient resolution, and because the density profile is not well-determined. With the advent of the Atacama Large Millimeter Array (ALMA), the estimated increase in angular resolution, to $\sim 0.1''$ \citep{Wooten2003}, should allow us to observe the form of these abundance profiles, so as to better determine  the initial flux-spectrum for cosmic rays for the Horsehead Nebula. 

It appears, from \citet{Indriolo2010}, that there is some environmental influence on the low energy flux of cosmic rays. It would be of great interest to not only examine the Horsehead Nebula at greater angular resolution, but to also observe and model other PDR's such as the Orion Bar, IC-63, L1688-W, and portions of Sgr B2 to determine how the low energy cosmic ray flux varies in our Galaxy. 
Sgr B2 is of special interest given the high values of $\zeta$ inferred from H$_3^+$ observations in this region \citep{Oka2005}.  Given the strong dependence of $\zeta$ on the path cosmic rays travel, it is very likely that the low-energy cosmic ray flux will be object-dependent.

\begin{acknowledgements}
We would like to thank Tom Millar, Ben McCall, and Nick Indriolo for helpful discussions about both the cosmic ray ionization rate and its implementation in chemical models. We also thank Ewine van Dishoeck, John Black, Tom Cravens and Andy Skilling for helping us properly consider the magnetic field.  E. H. acknowledges the support 
of the National Science Foundation for his astrochemistry program through grant AST-0702876, and his program in chemical kinetics through the Center for the 
Chemistry of the Universe. He also acknowledges support from NASA NAI for studies in the evolution of pre-planetary matter. E. R. acknowledges the support of the Observatoire de Paris and the Programme National du CNRS-INSU ``PCMI.''

\end{acknowledgements}

\bibliographystyle{aa}


\end{document}